\documentclass{emulateapj}

\slugcomment{To appear in the Astrophysical Journal}
\shorttitle{Extending the Model of KH 15D}
\shortauthors{Silvia \& Agol}

\begin{document}

\title{Extending the model of KH 15D: Estimating the effects of forward scattering and curvature of the occulting ring edge}

\author{
Devin W.\ Silvia\altaffilmark{1,2} and
Eric Agol\altaffilmark{1}
}

\altaffiltext{1}{Department of Astronomy, University of Washington, Box 351580, Seattle, WA 98105; dws129@astro.washington.edu, agol@astro.washington.edu}

\altaffiltext{2}{{\it Current address}: Department of Astrophysical and Planetary Sciences, University of Colorado, UCB 391, Boulder, CO 80309; devin.silvia@colorado.edu}

\begin{abstract}
The periodic eclipses of the pre-main-sequence binary, KH 15D, have been explained by a circumbinary dust ring inclined to the orbital plane, which causes occultations of the stars as they pass behind the ring edge.  We compute the extinction and forward scattering of light by the edge of the dust ring to explain (1) the gradual slope directly preceding total eclipse, (2) the gradual decline at the end of ingress, and (3) the slight rise in flux at mid-eclipse.  The size of the forward scattering halo indicates that the dust grains have a radius of $a \sim$6 (D/3 {\rm AU}) $\mu$m, where $D$ is the distance of the edge of the ring from the system barycenter. This dust size estimate agrees well with estimates of the dust grain size from polarimetry, adding to the evidence that the ring lies at several AU.  Finally, the ratio of the fluxes inside and outside eclipse independently indicates that the ring lies at a few astronomical units.
\end{abstract}


\section{Introduction}

The object KH 15D (V582 Mon) is a binary weak-lined T Tauri star in the cluster NGC 2264.  As a class, T Tauri stars are thought to be young stars that are still accreting gas from the remains of their parent molecular clouds.  This period of stellar evolution lasts for a few million years and is characterized by optical variability and chromospheric emission lines (see Bertout 1989 for an extended discussion).

Kearns \& Herbst (1998) reported a periodic flux variation in KH 15D:  it drops by 3.5 magnitudes every 48 days.  Such large flux variations ($>$2 mag in V) are seen in ``Type III" irregular variables in young clusters (Joy 1945, Herbst et al.\ 1994), but the remarkable property of KH 15D is that is {\it also} periodic.  This depth of eclipse can not be explained solely as a binary due to the large magnitude and duration, so the earliest theories of KH 15D postulated that circumstellar material eclipsed the star.

Since Kearns \& Herbst (1998) first reported the unusual properties of KH 15D, theoretical explanations for its properties have included an edge-on circumstellar disk (Hamilton et al.\ 2001, Agol et al.\ 2003, Winn et al.\ 2003), an asymmetric surrounding envelope (Grinin \& Tambovtseva 2002), and an orbiting vortex of solid particles (Barge \& Viton 2003).  However, it was soon realized that KH 15D was likely an eccentric binary system that is being occulted by the edge of a circumbinary dust ring (Winn et al.\ 2004, Chiang \& Murray-Clay 2004), and currently only one star is visible during part of its orbit causing the large change in magnitude.  In addition, the advance of the screen as a function of time due to the nodal precession of the ring, which is inclined relative to the binary, explains the lengthening of the duration of the eclipses as a larger portion of the orbit of the visible star is covered.  This model provides an excellent fit to the characteristics of the KH 15D light curve and radial velocities, as shown by Winn et al.\ (2004) and Chiang \& Murray-Clay (2004).  Chiang \& Murray-Clay (2004) also suggested that the ring must be somewhat warped in order to maintain the nodal precession.  This picture proved correct when KH 15D was confirmed to be an eccentric spectroscopic binary with a two-year, high-resolution, multi-site spectroscopic study (Johnson et al.\ 2004).  Archival data showed evidence for the second star which is completely hidden in recent data (Johnson et al.\ 2005).

Winn et al.\ (2006) made a further refinement of the KH 15D model by adding a faint blue halo around each star within the binary providing an explanation for the gradual slope of the light curve directly before and just after total eclipse, as well as the slight increase in flux during mid-eclipse.  In their model, they solved for the one dimensional halo brightness with 4 free parameters and an arbitrary functional form that resulted in a strongly asymmetric shape around both stars due to the asymmetric shape of ingress and egress, resulting in a puzzling physical picture: why should identical asymmetric halos surround both stars?  Winn et al.\ (2006) also included a gradual change in the angle of the edge of the ring projected onto the sky to account for longer-term variations the the lightcurve. 

In this paper, we focus on fitting the photometric data for KH 15D spanning 1995 to 2004 and the 18 published radial velocity data points.  We compute the effects of (1) a gradual change in opacity at the ring edge (prior models treat the ring edge as being sharp) and (2) forward scattering by dust at the ring edge, which together naturally explain the asymmetric shape of ingress and egress (Winn et al.\ 2006 acknowledge that forward scattering may be a more sensible explanation for their halo). We do not include the rotation of the edge of the ring (Winn et al.\ 2006), but instead propose that the edge of the ring is curved as a means of explaining additional variation in the eclipse durations that is not explained by the precession of the ring.  We show that these modifications provide a much better fit to the photometric data of KH 15D and we speculate on what additional insight this provides us about young stellar evolution and protoplanetary processes.  In section \ref{data} we summarize the data which we use to fit our model.  In section \ref{model} we discuss the elements included in our model.  In section \ref{results} we discuss the best-fit model parameters and errors, and in section \ref{summary} we discuss the implications of these results and possible future directions.

\section{Data}\label{data}

\subsection{Photometry}
We used the 6694 I-band photometric data points from Hamilton et al.\ (2005) taken between 1995 to 2004 with a dozen different telescopes using CCDs.  We ignored older, sparser data which is not precise nor dense enough to show the effects of forward scattering.  We converted the I-band magnitudes to fluxes using $f_\nu$/$f_0=10^{(-0.4 I)}$ where $f_0 \sim 2600$ Jy is the flux zero-point.  We enlarged the error bars for all data points in which the star was completely eclipsed or completely uneclipsed due to observed fluctuations that are larger than the reported errors.  Although these fluctuations may be partly due to starspots (Hamilton et al.\ 2005), for the purposes of model fitting we treat these fluctuations as an additional source of systematic error and replaced the errors bars that accompanied the data with the standard deviation of the scatter of the data in or out of eclipse, respectively. 

In addition to the orbital elements of the binary and the parameters describing the ring (described below), we introduce three primary flux parameters of our model: $\{f_{in}$, $f_{1,out}$, $f_{2,out}\}$, where $f_{in}$ is the non-variable flux of the system, $f_{1,out}$ is the flux received when only Star 1 is visible, and $f_{2,out}$ is the flux received when only Star 2 is visible.  

\subsection{Radial Velocities}
We fit our model to the published radial velocity data from Hamilton et al.\ (2003) and Johnson et al.\ (2004).  We did not discard data points taken near eclipse as these data points did not appear to show the Rosssiter-McLaughlin effect (Rossiter 1924, McLaughlin 1924), so we use all 18 values for the data fitting process.  We inflate the error bars for these radial velocity values to account for a systematic radial velocity offset due to light from the binary that is scattered off the back and sides of the circumbinary ring, an effect discussed by Herbst et al.\ (2008). If we assume that the parameter $f_{in}$ is due to the large angle scattering by the dust ring, which will have a range of Doppler shifts between $-K$ and $K$, where $K$ is the velocity semi-amplitude, then the systematic offset in the measured radial velocity values should be of order $\sqrt{2} \left({f_{in}/f_{1,out}}\right) K$.  This offset was added in quadrature to the uncertainties published by Winn et al.\ (2006).  Our decision to modify the error bars was to reduce the discrepancy between the orbital parameters found from the radial velocity data versus those found from the light curve data, as pointed out by Winn et al.\ (2006).  We implement weighting of the contribution to $\chi^2$ from photometric and radial velocity in the same manner as Winn et al.\ (2006), described in more detail in $\S$\ \ref{results}. 
\section{The KH 15D Model}\label{model}

\subsection{Orbit Model}
The binary orbit was computed by solving Kepler's equation, and converting the orbital elements to Cartesian coordinates and velocities, following the procedures described in Murray and Dermott (1999).  The parameters describing the binary orbit are $\{P, e, i, \omega, T_p, \gamma \}$, where $P$ is the period, $e$ is the eccentricity, $i$ is the inclination of the system, $\omega$ is the argument of pericenter, $T_p$ is the time of pericenter passage, and $\gamma$ is the heliocentric radial velocity of the center of mass.  The geometry of the orbit is shown in Figures \ref{fig1} and \ref{fig2}, where the origin is the barycenter of the binary, the Z-direction is along the line of sight, and we fix the longitude of the ascending node at $180^{\circ}$ so that the orbits cross the sky plane on the X-axis, with the visible star (which we call Star 1) crossing the sky plane on the positive X-axis as it moves towards positive Z.  We initially fit just the radial velocity data, confirming that our derived model parameters agreed with Winn et al.\ (2004).

\subsection{Ring Orientation, Occultation, and Curvature}
\begin{figure}
\plotone{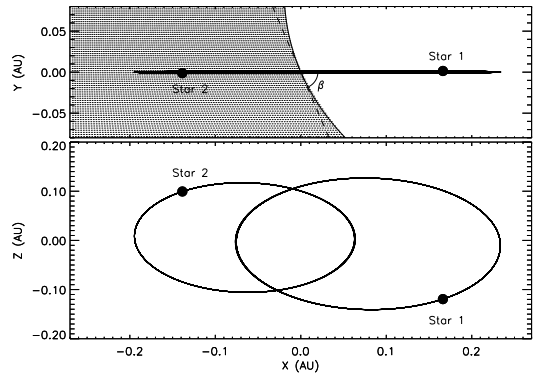}
\caption{TOP - The nearly edge-on orbits of the two stars in KH 15D as would be seen from the observer's perspective.  The solid curved line represents the edge of the occulting screen when the curvature parameter is added to the model while the dashed line represents the edge of the occulting screen with zero curvature.  $\beta$ is the angle between the X-axis and the edge of the screen as described in the text.  BOTTOM - The orbits of KH 15D as they would be seen from above.  In this representation, the occulting screen would not be visible.}
\label{fig1}
\vspace{0.1in}
\end{figure}

\begin{figure}
\plotone{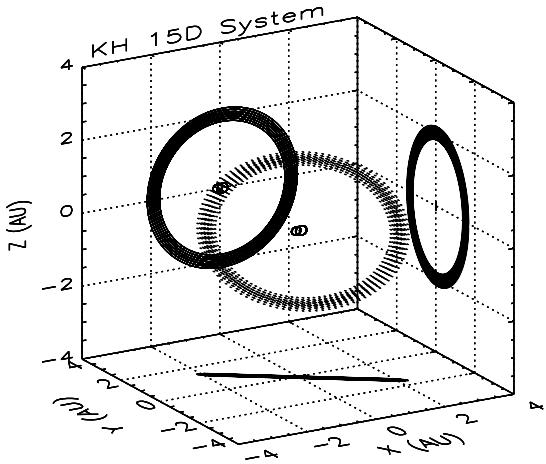}
\caption{A three dimensional representation of the geometry of the KH 15D system.  Although the orbits are defined by the best fit model, the distance between the barycenter of the system and the circumbinary ring is not fully known and so the plotted distance is just one possible value.  The shape and extent of the ring is not derived from our model, but is only shown to indicate one possible geometry.}
\vspace{0.1in}
\label{fig2}
\end{figure}

We define the angle between the ring edge, projected onto the $X-Y$ plane, and the X-axis as $\beta$.  We allow the edge of the ring to move in the Y-direction with velocity $v_{ring}$ and define the time at which the edge passes the center of mass of the system as $t_{ring}$.  Figure \ref{fig1} shows the orbits as they would be seen from the observer's perspective and a snapshot of the position of the edge of the ring.  To help visualize KH 15D, the approximate three-dimensional geometry is of the system is shown in Figure \ref{fig2}.

We assume the stars are limb-darkened with a quadratic limb-darkening law,
\begin{eqnarray}
I(r)&=&1-\gamma_1 (1-\epsilon) -\gamma_2\left(1-\epsilon\right)^2,
\end{eqnarray}
where $\epsilon$ is the cosine of the angle between the normal to the stellar surface and the line of sight to the observer.  We set $\gamma_1=0.4478$ and $\gamma_2=0.2091$, appropriate for a star with $T_{eff}$ $\approx$ 4300K (Agol et al.\ 2004) in the I-band according to the tables of Claret (2001).  

\begin{figure}
\plotone{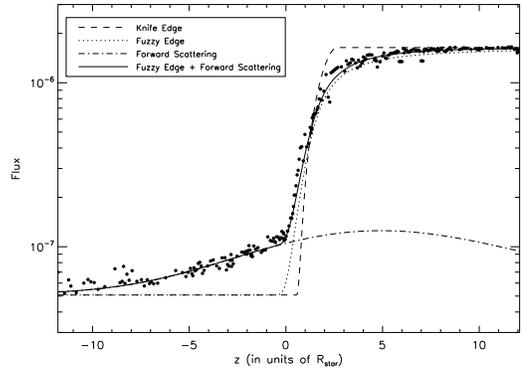}
\caption{The occultation curve for the currently visible star as a function of the distance between the center of the star and the point on the edge of the occulting screen where the optical depth becomes infinite.  The dots represent the data, but that data has been binned to allow for a clearer view of the agreement between the model and the observed values.  The solid line represents the full model while each of the other lines represents the shape of the curve due to different parts or versions of the model.  The shape of the forward scattering curve is directly dependent on model parameters; the height of the curve is determined by $\kappa_{as}$ while the width of the curve is determined by $w$.}
\label{fig3}
\vspace{0.1in}
\end{figure}

Initially we treat the ring as a ``knife edge'' which changes abruptly from transparent to opaque, and thus any part of the binary system seen above the edge is unobscured, while below the edge is completely occulted.  
For this sharp edge model, the flux during ingress and egress is given by
\begin{eqnarray}
F(z)&=&1+\langle I\rangle^{-1}\left[z\sqrt{1-z^2}\left(1-\gamma_1-
\frac{11}{6}\gamma_2 + \frac{1}{3}\gamma_2 z^2 \right)  \right.\cr
&&\left.- 
\left(1-\gamma_1-\frac{3}{2} \gamma_2\right) \cos^{-1}{z}\right.\cr
&&\left.- \frac{1}{6}\pi \left(\gamma_1 + 2 \gamma_2\right) \left(2 - 3z + 
z^3\right)\right],
\end{eqnarray}
where $z$ is the distance between the center of either star and the edge of the occulting ring in the units of the stellar radius and $\langle I\rangle=\pi (1-\frac{1}{3}\gamma_1-\frac{1}{6}\gamma_2)$.  This equation is only valid for $|z| < 1$ and takes on $F(z)=0$ for $z \le -1$ and $F(z)=1$ for $z \ge 1$.  Although the knife-edge model provides a fairly accurate set of initial parameters, the quantitative agreement with the data is poor, giving a $\chi^2$ of $\sim$12991 for 6697 degrees of freedom, where 6697 = 6694 photometric points + 18 velocity points - 15 free parameters. Most of the discrepancy in this fit occurs at the beginning of ingress and end of egress where the model has a much steeper slope than the data as shown in Figure \ref{fig3}.

The next step in our calculation is to add ``fuzziness'' to the edge of the ring by choosing the optical depth to vary as a power-law as a function of the sky-plane distance from the ring edge (it has infinite optical depth below the edge).  This power-law behavior seems physically plausible if we assume that there is a roll-off in density at the edge of the ring or due to the vertical structure of the ring.  We compute the transmitted flux during ingress and egress by convolving the knife-edge with the power-law optical depth,
\begin{eqnarray}
&&F^* =
{\alpha  \over x_0}\int^{max(0,z+1)}_{max(0,z-1)} dx e^{-(x_0/x)^\alpha} 
\left({x_0 \over x}\right)^{1+\alpha} F(z-x), \cr
&&
\end{eqnarray}
where $x_0$ is the scale length of the fuzzy edge (projected on the sky) in units of stellar radius, $R_{star}$ (we take both stars to have the same radius); $\alpha$ is the power-law exponent of optical depth variation; and $x$ is perpendicular to the ring edge.  In this model, we initially used a fixed value of $\alpha=2$ but after establishing a reasonable fit to the data we allowed $\alpha$ to vary freely.  We report $w$, which is $x_0$ transformed to physical units, $w\equiv x_0 R_{star}$ (in meters). This addition reduces $\chi^2$ to 12890 for 6695 degrees of freedom, improving the fit by $\Delta \chi^2 = 101$.

We next add curvature of the ring edge to the model which improves the fit to mid-eclipse data.  We parameterize this as $y_{ring} = \mu x_{ring}^2$, where $y_{ring}$ and $x_{ring}$ are the $y$ and $x$ coordinates perpendicular and parallel to the edge of the ring at the origin when it crosses the barycenter of the system.  We find that by adding curvature the rate at which the central re-brightenings fade during 1995-1998 is slower than without it since the hidden star's orbit remains visible for a slightly longer period of time, agreeing better with the lightcurve data.  Again we see an incremental improvement in the model as $\chi^2$ goes to 9582 with 6694 degrees of freedom, producing $\Delta \chi^2 = 3308$.  We neglected precession of the ring as the model of Winn et al.\ (2006) indicates the edge projected on the sky plane would have rotated only $\sim$4.5 degrees over the 9 year data set.  

\subsection{Forward Scattering}

Forward scattering by dust is the most important new addition to our model: the same dust at the edge of the ring causing extinction will also diffract light, leading to an apparent halo around the star.  The same effect can be seen on nights when the moon passes behind a thin layer of clouds.  The water droplets in the clouds diffract light from the moon producing a halo of forward-scattered light about the moon.  For both the moon and KH 15D the halo is present when the optical depth is near unity --- at very high optical depths multiple scatterings will cause the radiation to eventually be absorbed, while at very low optical depths there is almost no scattering.  As the star passes behind the edge of the ring, this forward scattering halo softens the shape of the ingress and egress, as observed in KH 15D.

The angular distribution of scattered light for spherical grains can be modeled approximately as an Airy disk.  Since the wings of the Airy disk are much weaker than the central peak, we approximate the Airy disk by a Gaussian angular distribution to allow for faster computation of the scattered flux,
\begin{equation}
{d \sigma_F \over d \Omega } = \sigma_0 e^{-{\theta^2\over 2\sigma_\theta^2}},
\end{equation}
where $d\sigma_F/d\Omega$ is the differential scattering cross section into
a solid angle $d\Omega$.
A Gaussian with a standard deviation of $\sigma_\theta = 0.43 \lambda/(2\pi a)$, where  $a$ is the radius of the scattering particle, gives a good approximation to the peak of an Airy disk.  There may be additional large-angle scattering by the ring; however, if the size of the ring is much larger than the orbital size of the binary, then this radiation should have a much weaker time dependence than the forward-scattered light, so we treat this as a constant flux contribution during the eclipse, namely $f_{in}$ mentioned in $\S$\ 2.1. 

As the optical depth becomes large, multiple forward scatterings can occur.  We make the assumption that the scattering region (due to the ring surrounding the two stars) can be treated in the thin-screen approximation; that is, (1) the scattering angle is small enough that the photon can be approximated as going straight as it passes through the dust ring, $\cos{(\theta_{scat})} \sim 1$, and thus the total optical depth for each forward-scattered photon is simply the line-of-sight optical depth to the star; and (2) the dust is at approximately a constant distance from the source.  Since absorption and large-angle scattering can occur in addition to forward scattering, we add an additional parameter, $\kappa_{as}$, the ratio of the sum of the absorption and large-angle scattering opacity to the forward scattering opacity.

With these approximations, we can compute the specific intensity of scattered radiation, $I(x,y)=\sum_{N=1}^{N_{max}} I_N(x,y)$ at a position $(x,y)$ relative to the source position
$(x_0,y_0)$, both positions projected on the sky plane and relative to the ring edge (which is rotated relative to the $X-Y$ plane).  For radiation that has undergone $N$ scatterings, the specific intensity of the scattered radiation, $I_N(x,y)$, is given by a Gaussian, with standard deviation $\sqrt{N}\sigma_{\theta}$, times the probability of scattering $N$ times,
\begin{eqnarray}
&&I_N (x,y) = \cr
&&\ \ \ \ \ \ \ \ \  {F_0 \over 2\pi N \sigma_\theta^2} 
{\tau^N(x,y)\over N!} e^{-\tau(x,y)(1+\kappa_{as})}
e^{-{(x-x_0)^2+(y-y_0)^2 \over 2 D^2 N 
\sigma_\theta^2}}, \cr
&&
\end{eqnarray}
where $F_0$ is the flux of the star at the distance of the dust scattering screen, $D$ is the star-dust separation, and $\tau$ is the  forward-scattering optical depth through the slab at position $(x,y)$.  We treat the stars as point sources for the purposes of computing the forward-scattered light.   When $\kappa_{as} \sim 1$, truncating the sum at $N_{max}=5$ scatterings gives a converged distribution of scattered light.  We report $\sigma_{D} \equiv D\times\sigma_\theta$ in meters rather than $\sigma_\theta$.

\begin{figure}
\plotone{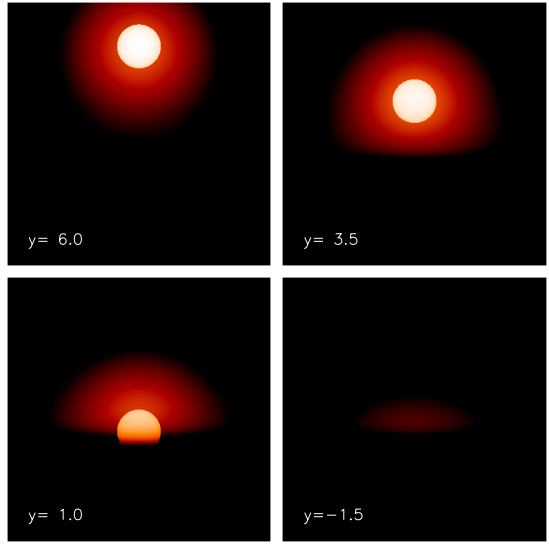}
\caption{Star and forward scattering halo as a function of the height above the edge of the ring. Each frame is 12 stellar radii on a side, and $y$ represents the height of the center of the star above the edge of the ring (where $\tau=\infty$) in units of the stellar radius.  The grey (or color) scale represents the logarithmic I-band intensity where white is the surface brightness of the center of the star while black is $\le 10^{-4}$ of this brightness. Limb-darkening of the star is included, but the star appears nearly uniform due to the logarithmic intensity scale.  An mpeg version of this figure is available on-line.}
\label{fig4}
\vspace{0.1in}
\end{figure}

Figure \ref{fig4} shows the distribution of scattered light as a function of distance of the center of the star from the edge ($\tau=\infty$) of the ring.   The best-fit parameters (see below) for the KH 15D system were used in creating this figure.  When the star is far above the ring edge ($y=6.0$ frame), the forward scattering is weak as the optical depth is small within the angular size of the scattering halo.  As the star approaches the edge, forward scattering increases as the optical depth increases ($y=3.5,1.0$ frames), but the stellar light starts to decrease due to extinction by the dust.  As the star becomes eclipsed, the halo can still be seen above the edge, but it is much weaker due to extinction of the scattered halo ($y=-1.5$ frame).

For computing the sky-plane separation of the star and ring edge, we take into account the curvature of the edge of the ring since the distance traveled by the star and the edge of the ring is comparable to the radius of curvature.  However, to simplify the scattering calculation we treat the edge as being straight (since the radius of curvature is much larger than the edge scale-length) and the optical depth varies only perpendicular to the ring edge, allowing analytic integration of the scattered specific intensity parallel to the edge (along $x$), while integration perpendicular to the edge is carried out numerically.  

After adding forward scattering, the model light curve matches the more gradual slope on either end of the eclipses as well as the small rises in flux during mid-eclipse.  The $\chi^2$ reduces to 7162 for 6692 degrees of freedom, an overall improvement of $\Delta \chi^2 = 5829$ as compared to the knife-edge model with no forward-scattering.  In particular, we notice that the combination of forward scattering with a curved ring edge produces mid-eclipse ``bumps" in the model that are more pronounced and widened, which agrees well with the data. Furthermore, we see the noticeable effects of the hidden star decrease as the heights of the mid-eclipse bumps diminish over time (Figure \ref{fig3}).  This supports the idea that as time passes an increasing portion of the orbit of the binary system is obscured by the ring of material.

As a comparison between our model and the ``halo" model of Winn et al.\ (2006), we take the flux data as a function of the separation between Star 1 and the edge of the ring (as in Figure \ref{fig3}) and run a $\chi^2$ minimization using the halo model on the data points for which Star 2 contributes an insignificant flux.  Using the seven parameters necessary for modeling the occultation curve, $\{R_{star}, f_{in}, f_{1,out}, w, \sigma_{D}, \kappa_{as}, \alpha \}$ for our model versus $\{R_{star}, f_{in}, f_{1,out}, \epsilon_1, \epsilon_2, \xi_1, \xi_2 \}$ for Winn et al.'s (2006) model, we find that forward scattering produces $\chi^2 = 5849.4$ and the halo model produces $\chi^2 = 5903.7$ for 5484 degrees of freedom. The difference in $\chi^2$ between the two models is  $\Delta \chi^2 = 54$, indicating that the forward scattering model is favored over the halo model.

\section{Results}\label{results}

The complete model has 18 free parameters, consisting of the 6 orbital elements, $\{P$, $e$, $i$, $\omega$, $T_p$, $\gamma \}$, 4 parameters defining the orientation, shape, and motion of the occulting screen, $\{v_{ring}$, $t_{ring}$, $\beta$, $\mu \}$, and 8 parameters that modify the structure of the lightcurve, $\{R_{star}$, $f_{in}$, $f_{1,out}$, $f_{2,out}$, $w$, $\sigma_{D}$, $\kappa_{as}$, $\alpha \}$.  In addition, we assume the masses of the stars are $M_1=0.6 \pm 0.1$ M$_{\odot}$ and $M_2=0.72 \pm 0.1$ M$_{\odot}$, respectively.  This is based on the theoretical pre-main-sequence evolutionary tracks and the mass-luminosity relation indicating that the mass of the visible star should be 0.6 $\pm$ 0.1 M$_{\odot}$ and the mass ratio, $M_2$/$M_1$, should be 1.2 $\pm$ 0.1 (Winn et al.\ 2006).

\begin{figure*}
\plotone{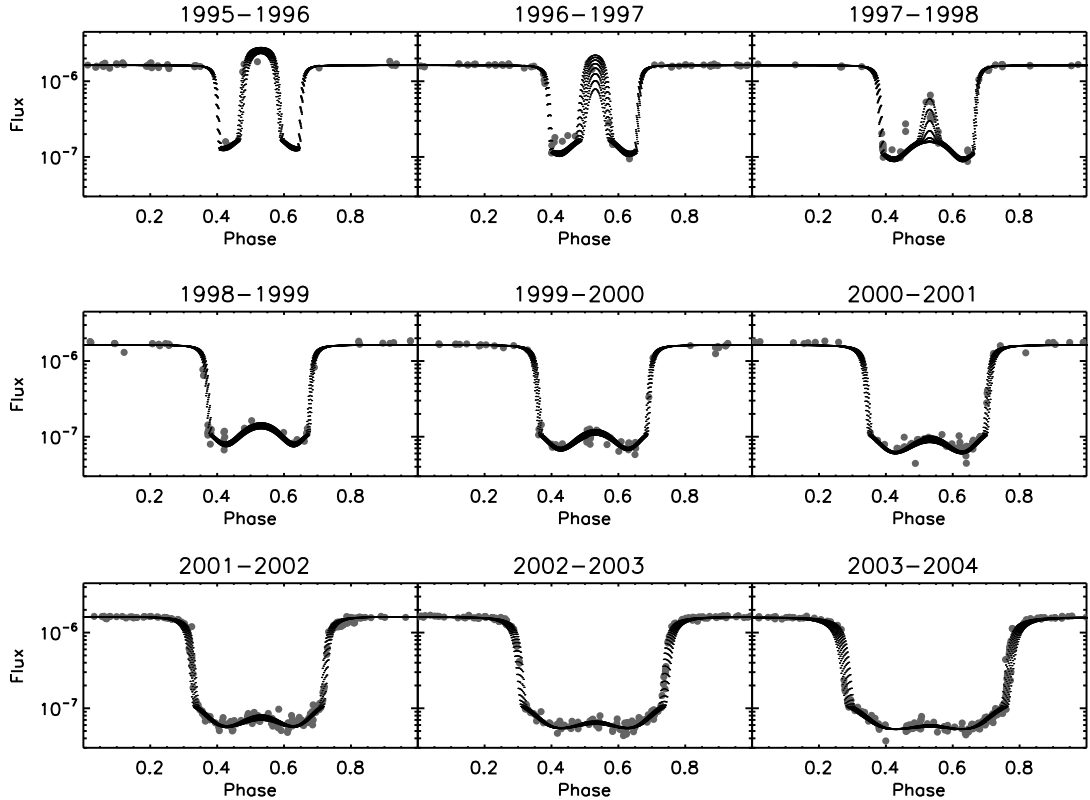}
\caption{The lightcurve of KH 15D as a function of phase.  Each plot shows one year worth of data overplotted with the best fit model.  The most notable features are the drop off in mid-eclipse rebrightenings from 1995-1998, the gradual deepening of the eclipses from start to finish, and the small rises in flux during mid-eclipse from 1998-2004, which are produced by forward scattering in our model as the ring edge moves across the orbit of the binary.} 
\label{fig5}
\end{figure*}

\begin{figure}
\plotone{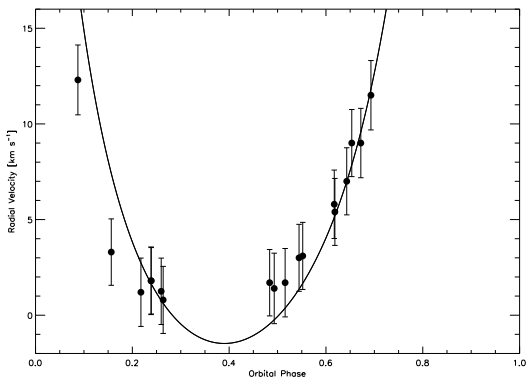}
\caption{The radial velocity profile of KH 15D as a function of phase.  The solid line represents the model while the dots represent the 18 measured radial velocities and their associated errors bars as defined in the text.}
\label{fig6}
\vspace{0.1in}
\end{figure}

For the optimization of these parameters we choose to minimize with respect to a $\chi^2$ defined in a similar way to that of Winn et al.\ (2006), such that
\begin{equation}
\chi^2 = \sum_{i=1}^{N_f} \left({{f_{mod,i} - f_{data,i}} \over \sigma_{f,i}}\right)^2 + \lambda \sum_{i=1}^{N_v} \left({{v_{mod,i} - v_{data,i}} \over \sigma_{v,i}}\right)^2,
\end{equation}
where $N_f = 6694$ and $N_v = 18$.  The presence of $\lambda$ allows us to apply a certain amount of weight to the relatively few radial velocity measurements so that equally good fits are found to both fluxes and radial velocities.  To determine what value of $\lambda$ to use for our model, we attempted to find a value that would produce $\frac{\chi_{f}^2}{N_{f}} \approx \frac{\chi_{v}^2}{N_{v}}$, where $\chi_{f}^2$ and $\chi_{v}^2$ are the separate non-reduced $\chi^2$ values for the fluxes and radial velocities respectively. This procedure gives $\lambda = 9$ for our final model.  Using that value of lambda we find that our best-fit parameter set produces $\chi_{f}^2 = 6992.6$ and $\chi_{v}^2 = 18.8$.  A list of the best-fit parameters and errors (described next) is presented in Table \ref{tab1}.  In addition, Figures \ref{fig5} and \ref{fig6} show the model and corresponding data for both the lightcurve and the radial velocities.

\begin{deluxetable}{cccrrrrrrrrrrrrrrrrrrrr}
\tabletypesize{\normalsize}
\tablecaption{Model Parameters for KH 15D \label{tab1}}
\tablewidth{0pt}
\tablehead{\colhead{Parameter} & \colhead{Estimated Value} & \colhead{Units}}
\startdata
$M_1$ & 0.6 $\pm$ 0.1 & $M_{\odot}$ \\
$M_2$ & 0.72 $\pm$ 0.1 & $M_{\odot}$ \\
$P$ & 48.359 $\pm$ 0.0012 & days \\
$e$ & 0.51 $\pm$ 0.008 & --- \\
$i$ & 89.31 $\pm$ 0.38 & degrees \\
$\omega$ & 5.1 $\pm$ 1.3 & degrees \\
$T_p$ & 2352.72 $\pm$ 0.13 & J.D.-2450000 \\
$\gamma$ & 18.6 $\pm$ 1.5 & km s$^{-1}$ \\
$v_{ring}$ & 58.46 $\pm$ 5.95 & m s$^{-1}$ \\
$t_{ring}$ & -1289.64 $\pm$ 340.23 & J.D.-2450000 \\
$R_{star}$ & 1.2 $\pm$ 0.23 & $R_{\odot}$ \\
$f_{in}$ &  $(5.075 \pm 0.123)\times 10^{-8}$ & $f_0$ \\
$f_{1,out}$ & $(1.59 \pm 0.03)\times 10^{-6}$ & $f_0$ \\
$f_{2,out}$ & $(3.28 \pm 0.28)\times 10^{-6}$ & $f_0$ \\
$w$ & $(1.30\pm 0.67)\times 10^9$  & m \\
$\sigma_{D}$ & $(4.35 \pm 0.34)\times 10^9$ & m \\
$\kappa_{as}$ & 1.91 $\pm$ 0.18 & --- \\
$\beta$ & 1.18 $\pm$ 0.09 & radians \\
$\mu$ & 2.00 $\pm$ 0.40 & AU$ ^{-1}$ \\
$\alpha$ & 1.53 $\pm$ 0.59 & --- \\
\enddata
\end{deluxetable}

To estimate the uncertainties in each of our 18 parameters we followed the procedure of Winn et al.\ (2006).  We repeatedly fit our model to artificial data sets with random noise added to our best-fit model.  To model the noise, we took the residuals of the best-fit model, normalized them by the model flux, randomized them, and then added them back to the best-fit model (again scaling to the flux) to create an artificial data set.  Since the number of radial velocities points is so small, we instead added Gaussian noise with a standard deviation equal to the error bars of the radial velocity data (Winn et al.\ 2006).  For each artificial data set we also allowed $M_1$ and $M_2$ to vary with Gaussian distributions that had means of 0.6 and 0.72 respectively and standard deviations of 0.1 so that the resulting error bars would have the $\pm$ 0.1 $M_{\odot}$ constraint placed on them as mentioned above.  To determine the error bars on the remaining parameters, which are quoted in Table \ref{tab1}, we took the standard deviations of ~2000 parameter sets that were produced from fits to 2000 random realizations of the noisy lightcurve.  We found that subsequent iterations did not appear to produce significant changes in the errors.

\section{Summary and Discussion}\label{summary}

We find that our best-fit model is in good agreement with the data with a reduced $\chi^2$ of order unity.  The model accurately reproduces the fall-off in the emergence of Star 2 during the 1995-1998 time frame as well as the gradual deepening of the total eclipses from 1998-2004 (Figure \ref{fig5}).  Furthermore, we find one of the most satisfying results to be the success of forward scattering as the source for the mid-eclipse bumps in the lightcurve once Star 2 no longer passes beyond the edge of the ring, as well as explaining the gradual fall off and rises in the curve as Star 1 enters and exits the eclipses (Figure \ref{fig3}).  Our model also predicts that around the beginning of 2008 the photosphere of Star 1 will no longer move beyond the edge of the occulting screen -- the same time frame as predicted by Winn et al.\ (2006).  After that point, the maximum light we will receive from KH 15D will be due to forward scattering as Star 1 nears the edge of the ring, which will last roughly 5-7 years.

Several of the predicted parameters are in relatively good agreement with the values we expect.  First, $\gamma$ is in agreement with both the value published by Winn et al.\ (2006) as well as the median heliocentric radial velocity of the cluster NGC 2264, 20 $\pm$ 3 km s$^{-1}$ (Soderblom et al.\ 1999).  Second, we allowed the radius of Star 1, $R_{star}$, to be a free parameter determined by the shape of the lightcurve, finding a best-fit of 1.2$\pm 0.23 R_\odot$ which is consistent with the value used in Winn et al.\ (2006), who held the parameter fixed at 1.3 $R_{\odot}$ based on the mass-radius relation as predicted by pre-main-sequence evolutionary tracks.  Finally, we find that our value for the eccentricity, 0.51, is consistent with the pseudo-synchronization {\it upper} limit of 0.66 as discussed by Winn et al.\ (2006).

Our derived value for the angle between the edge of the ring and the horizontal axis of the system, $\beta=68^\circ \pm 5^\circ$, was not what we expected.  Although our error bars seem to indicate that the value is relatively well-constrained, since the system is nearly edge on, changes in $\beta$ will not produce significant changes in the model lightcurve, so we are not confident in the best-fit value or uncertainty of this parameter.

We now compare our best-fit model with the model in Winn et al.\ (2006).  Our best-fit period, eccentricity, and inclination are relatively close to those presented by Winn et al.\ (2006), agreeing to within 0.1\%, 10\%, and 5\%, respectively.  However, formally there is quantitative disagreement ($>1 \sigma$) between Winn et al.'s (2006) values and our own which most likely results from physical differences between their model and ours, as well as fitting different data sets with different assumed errors.   Our best-fit flux for Star 2 is nearly twice that of Star 1, somewhat larger than the ratio of 1.36 found by Winn et al.\ (2006).

From the forward scattering width in our model, $\sigma_\theta$, we can estimate the average size of the dust grains which dominate the forward-scattering opacity.  Using the parameter $\sigma_{D}$ and our Gaussian approximation $\sigma_\theta = 0.43 \lambda/(2\pi a)$ with $\lambda = 8140$  \AA, we find that the radius $a$ of the dust grains is $\sim$6(D/3 AU) $\mu$m, where D is the distance between the edge of the ring and the stars. This agrees extremely well with the  $\sim$6-8 $\mu$m size of the dust grains estimated by Agol et al.\ (2004) based on the weak variation of the observed polarization with wavelength (smaller/larger grains cause a stronger/weaker variation of polarization with wavelength than observed).  A distance of 3 AU was estimated by Chiang \& Murray-Clay (2004) and Winn et al.\ (2004) to explain a rate of precession that matches the observed velocity of the ring edge across the binary, indicating internal consistency between the forward-scattering, polarization, and precession constraints on $a$ and $D$.

Beyond our confirmation that forward scattering is the correct physical process to produce the lightcurve features, we can also use the flux parameters to estimate the apparent area of the ring.  One possible interpretation of the residual flux in eclipse is that it is due to large-angle scattering off of the ring.  This implies that the ratio of $f_{in}$ to the sum of $f_{1,out}$ and $f_{2,out}$ is approximately the ratio of scattered flux from the ring to the total flux of the stars, and thus gives an indication of the solid angle covered by the ring.  From our parameters, $\frac{f_{in}}{f_{1,out}+f_{2,out}} \approx 0.01$.  If we then make a simplifying assumption that the ring of occulting material has roughly a shape of a warped ring with mean radius $R$ and maximum warp angle $\psi$ (the difference between the inner and outer edges), then the solid angle covered by the warped ring is on average $\sim 2\pi\psi$.  Let us then say that perhaps we only see the back half of the warped ring, that only half of the light that hits the ring is reflected back towards us, and that of order half of the back of the ring is obscured by the front of the ring.  This would imply that $f_{in} \approx \frac{1}{2^3} \psi (f_{1,out}+f_{2,out})$.  Since we observe a value of $\frac{f_{in}}{f_{1,out}+f_{2,out}} \approx 0.01$, this indicates that $\psi \approx 0.08$.  We also know that the warped ring at the very least covers the orbit of the stars, and since our fit implies that the edge is at a large angle to steller orbits (which are nearly edge-one), which implies $R\psi > 0.3$ AU and thus $R > 0.3/0.08 \sim 4$ AU.  This argument contains several highly uncertain geometry-dependent and scattering-dependent factors, but is another consistency check, independent of the above arguments, that the ring lies at a few astronomical units.  Further modeling of the ring is definitely warranted.

Additional tests of our model can also be carried out to confirm its accuracy.  One such test would be to see if a scattering halo is present at infrared wavelengths.  At longer wavelengths, there will be competing effects between a larger forward scattering angle and a smaller absorption opacity.  We are currently applying our model to Spitzer data which will be published once data analysis is complete.  Another test would be to compare the radial velocities of the forward-scattered stellar light and the large-angle scattered light.  The forward-scattered light should have a radial velocity equal to that of the eclipsing star, while the large-angle scattered light should have a wide range of radial velocities as the light is scattered off of different parts of the ring (Herbst et al.\ 2008).

Our particular choice for the parameterization of the optical depth of the ring-edge as a power-law should be taken to be fairly generic since for the purposes of modeling the scattering halo we are only concerned with the opacity near $\tau \sim 0.1-3$ and over this small range of $\tau$ a power-law functional form should provide a good fit to almost any smooth variation.  Predictions for the variation in surface density near a gap caused by a planet (if the ring is shepherded by a planet) obey a more complex functional form, but depending on how the surface density maps to opacity in the I-band, only the small region near $\tau \sim 1$ can be approximated by a power law.  

The parameter $\mu$ is essentially an estimate of the sky-plane radius of curvature, $\mu \sim 1/(2R)$.  Our best-fit value of $\mu=2$ AU$^{-1}$ implies a sky-plane radius of curvature of 0.25 AU, which is a rather small value if the ring edge is located at $\sim$ 3 AU.  This might imply strong warping of the ring, or it may simply be that this is not the correct parameterization of the properties of the ring edge. 

Rice et al.\ (2006) produced simulations of a ring with an imbedded planet around a T-Tauri star.  They found that the surface density of their rings varies by an order of magnitude over a distance that is $\sim 0.1$ of the semi-major axis.  In our model the same variation occurs over $\sim 5 w \sim 6\times 10^{9}$ m, about 1/8 of the value predicted by the Rice et al.\ (2006) model, which agrees reasonably well if the ring is nearly edge-on causing a smaller sky-plane scale-length.  This is not at all evidence that a planet is present in the system, but a shepherding planet may be one explanation for the variation in optical at the edge of the ring.  The scale length should depend on opacity which varies with wavelength, an additional reason to make infrared observations.

To finish our discussion of model parameters, we comment on $\kappa_{as} \sim 2$.  As the wavelength of light is $\lambda=0.8$ $\mu$m and we have estimated the radius of the dust particles is $a = 6$ $\mu$m, then the particle scattering parameter is $x=2\pi(\frac{a}{\lambda}) \approx 50$.  In this limit, the forward scattering cross section and reflectance (large-angle scattering plus absorption) cross section equal the area of the particle, so one would expect $\kappa_{as} = 1$.  Therefore, our value of $\kappa_{as} \approx 2$ indicates that there may be grains smaller than 6 $\mu$m mixed in with $x \approx 1$.  These smaller grains would not have as narrow a forward-scattering peak and would thus contribute to the large-angle scattering/absorption component --- resulting in an increase in the value of $\kappa_{as}$. Since large-angle scattering causes polarization and forward scattering does not, a possible test of the model proposed in this paper would be to measure the degree of polarization as a function of the ratio of forward-scattered to large-angle scattered light, which changes during the ingress/egress.   Infrared, radial-velocity, and polarimetric measurements, combined with more detail modeling of the dust ring, should help to put tighter constraints on the properties of the system and unlock further mysteries of KH 15D.

\acknowledgements
We would like to thank the referee for helpful comments on the paper.  Partial support for this work was provided by NASA through an award issued by JPL/Caltech.  DS acknowledges support from a University of Washington Mary Gates scholarship and two NASA Space Grant Summer Undergraduate Research Program stipends.  EA acknowledges support from NSF CAREER grant Award 0645416.

\end{document}